\documentclass[sigconf]{acmart}
\renewcommand\footnotetextcopyrightpermission[1]{}

\acmConference[ ]{ }{ }

\AtBeginDocument{%
	}

\begin{document}

	%%\title{The Name of the Title Is Hope}
	\title{FastBCSD: Fast and Efficient Neural Network for Binary Code Similarity Detection}

	\author{Chensen Huang}
	\affiliation{%
		\institution{University of Chinese Academy of Sciences}
		\city{}
		\country{}}
	\email{}

	\author{Guibo Zhu}
	\affiliation{%
		\institution{Institute of Automation, Chinese of Academy}
		\city{}
		\country{}}
	\email{}
	
	\author{Guojing Ge}
	\affiliation{%
		\institution{Institute of Automation, Chinese of Academy}
		\city{}
		\country{}}
	\email{}
	
	\author{Taihao Li}
	\affiliation{%
		\institution{Zhejiang Lab}
		\city{}
		\country{}}
	\email{}

	\author{Jinqiao Wang}
	\affiliation{%
		\institution{Institute of Automation, Chinese of Academy}
		\city{}
		\country{}}
	\email{}
	
	\renewcommand{\shortauthors}{Huang et al.}
	
	\begin{abstract}
		Binary code similarity detection (BCSD) has various applications, including but not limited to vulnerability detection, plagiarism detection, and malware detection. Previous research efforts mainly focus on transforming binary code to assembly code strings using reverse compilation and then using pre-trained deep learning models with large parameters to obtain feature representation vector of binary code. While these models have proven to be effective in representing binary code, their large parameter size leads to considerable computational expenses during both training and inference. In this paper, we present a lightweight neural network, called FastBCSD, that employs a dynamic instruction vector encoding method and takes only assembly code as input feature to achieve comparable accuracy to the pre-training models while reducing the computational resources and time cost.
		On the BinaryCorp dataset, our method achieves a similar average MRR score to the state-of-the-art pre-training-based method (jTrans), while on the BinaryCorp 3M dataset, our method even outperforms the latest technology by 0.01. Notably, FastBCSD has a much smaller parameter size (13.4M) compared to jTrans (87.88M), and its latency time is 1/5 of jTrans on NVIDIA GTX 1080Ti.
		
	\end{abstract}
	
		
		
		%%  \includegraphics[width=\textwidth]{sampleteaser}
		%%  \caption{Seattle Mariners at Spring Training, 2010.}
		%%  \Description{Enjoying the baseball game from the third-base
			%%  seats. Ichiro Suzuki preparing to bat.}

		%%\received{20 February 2007}
		%%\received[revised]{12 March 2009}
		%%\received[accepted]{5 June 2009}
		
		\maketitle

		\section{INTRODUCTION}
		Binary code similarity detection (BCSD) is a technique used in computer security fields, such as vulnerability detection, clonal detection\cite{c24,c25}, and malware detection\cite{c26,c27}, that calculates the similarity between two binary code fragments. As binary analysis tasks become more widespread, there is a clear need to develop faster and more efficient BCSD solutions.
		
		Gemini \cite{c1} and Genius \cite{c2} are earlier works that manually extract statistical features from the instruction and combine them with control flow diagrams (CFG) to model this task. In contrast, $\alpha$Diff \cite{c3} directly feeds binary data into the neural network to extract representation features. As natural language processing (NLP) technology has developed, recent research has explored new approaches to binary code similarity detection (BCSD). One such approach is SAFE \cite{c13}, which treats the entire instruction (including both operand and opcode) as a word (token) and applies deep learning methods to learn a vector representation. Another approach is Instruction2Vec \cite{c21}, which splits operands and opcodes in the instruction into separate words (tokens), assigns each token a learnable vector, and trains a deep learning model on the resulting vector representations. In recent years, pre-training models have gained popularity in the field of binary code similarity detection (BCSD). PalmTree \cite{c10}, jTrans \cite{c11}, and UniASM \cite{c12} are three such models that use cleaned assembly text as a sequence of tokens, along with pre-training tasks like Masked Language Model (MLM)\cite{c8} and Context Window Prediction (CWP)\cite{c10} to conduct unsupervised training. These models have shown promising results in BCSD. The representation vector of the entire assembly text can be obtained by fine-tuning the assembly code pair. The pre-training approach has demonstrated significant performance gains over previous techniques, but there are still some limitations that need to be addressed.
		
		Firstly, the large number of parameters in recent models based on BERT and other Transformer structures have made it difficult to apply them in real-world scenarios.
		
		Secondly, the Transformer \cite{c7} has limitations that the length of input sequences cannot exceed 512 at a time. As a result, when the length of input sequences exceeds 512, some sequences will be discarded, resulting in a loss of information.
		
		Thirdly, in order to address the Out-Of-Vocabulary (OOV) problem, token normalization is commonly applied, which maps different tokens to the same token, but excessive normalization can result in information loss.
		
		In response to the aforementioned issues, this paper introduces a novel dynamic instruction vector encoding approach and utilizes a lightweight neural network model to extract instruction characteristics. In contrast to UniASM \cite{c12}, where each instruction consists of both operands and opcodes as a single token, we embed the operands and opcodes as separate tokens and concatenate the opcode vector, operand vector, and instruction position vector along the feature dimension. Additionally, the length of the dynamic instruction vector is proportional to the number of instructions, which is much smaller than the number of tokens, resulting in reduced training and inference time, as well as mitigating the risk of OOV issues. During token generation, various delimiters (e.g., ",", ":", ";", etc.) are employed to split the assembly code into multiple tokens. Low-frequency tokens are then filtered out based on their occurrence frequency. In contrast to normalizing tokens, this approach reduces the number of tokens while preserving more token information. we selected TextCNN \cite{c14} and LSTM \cite{c15}, which are well-established and widely-used models for text classification and other natural language processing tasks, as our basic model. The TextCNN model, which is based on one-dimensional convolution, has the advantages of having a small number of parameters and fast inference speed. On the other hand, the LSTM model based on RNN can easily extract the overall features of the token sequence. Recent studies have shown that the MLP-Mixer\cite{c19} model based on MLP achieved comparable performance to Transformer on a series of vision tasks\cite{c28} and language tasks\cite{c29}. Compared to the Transformer, the MLP-Mixer has a simpler architecture that utilizes only multi-layer perceptrons and lacks the incorporation of self-attention. We will modify the MLP-Mixer used in vision tasks to make it applicable to the BCSD task.
		
		In summary, our study offers the following contributions:
		\begin{itemize}
			\item FastBCSD is easy to follow compared to other research methods, as it only requires assembling text strings and a lighten neural network, while other methods necessitate additional features, numerous training techniques, and complex architecture.
			
			\item The performance of FastBCSD using TextCNN is comparable to the state-of-the-art (SOTA) model: jTrans \cite{c11} based on pre-trained models, but with a significantly reduced computational time of approximately 1/5 of jTrans and a smaller number of parameters, which is approximately 1/5 of jTrans.
		\end{itemize}

		\section{RELATED WORK}
		In prior research, researchers commonly employed direct analysis of specific features of binary code. For instance, $\alpha$Dif \cite{c3} used a CNN \cite{c17} model to extract internal features of each binary function, which processed raw bytes without additional feature requirements. Subsequent research schemes, however, typically focused on seeking a vector to represent binary code, with the identification of a suitable representation vector serving as the crux of these approaches. The authors of DeepVSA \cite{c18} employ One-hot encoding on the raw bytes to obtain the representation vector of each instruction, using it to classify malicious software. In contrast, the authors of Gemini \cite{c1} construct an attribute control flow graph (ACFG) by manually extracting statistical features of the assembly code, such as the number of constants. A graph embedding network is trained to generate representation vectors in this approach. The author of \cite{c20} has proposed a technique to represent binary code as a sequence of instructions, and then applied the word2vec algorithm \cite{c22} to obtain a vector representation for each instruction. A recurrent neural network based on LSTM \cite{c15} was then used to identify similar binary code. This method shares similarity with the approach adopted in SAFE \cite{c13}. Instruction2Vec \cite{c21} first pre-trained tokens for opcodes and operands using word2vec \cite{c22}, and represented each instruction as a vector that combines an opcode vector and eight operand vectors. The resulting vector size is N x 9 x vector-size, where N is the instruction length. The researchers then used a CNN \cite{c14} model for training. Asm2Vec \cite{c23} employs random walks on the CFG (Control Flow Graph) to sample instructions and uses a model similar to the PV-DM model to train function and instruction tokens to obtain representation vectors. Pre-trained models have achieved remarkable results in the field of NLP, and have also been applied to the BCSD task in recent years with favorable outcomes. The study by researchers in OrderMatters \cite{c9} proposes a semantic-aware neural network to extract semantic information from binary code. By pre-training binary code using BERT \cite{c8} at token, block, and graph level tasks, the researchers found that the order of Control Flow Graph (CFG) nodes plays a crucial role in detecting graph similarity. To address the issue of the importance of the order of CFG nodes in graph similarity detection, the authors of the OrderMatters study utilized a Convolutional Neural Network (CNN) to extract order information from the adjacency matrix. The extracted features were then integrated to form the final representation vector for binary code. PalmTree \cite{c10} is an assembly language model pre-trained using the BERT model architecture. In the study, the researchers performed self-supervised training on a vast unlabeled binary corpus to generate universal instruction embeddings. In the study of PalmTree \cite{c10}, researchers proposed a pre-training model based on BERT called PalmTree, which utilizes three pre-training tasks: Masked Language Model (MLM), Context Window Prediction (CWP), and Def-Use Prediction (DUP) to extract various features of assembly language. Although the model generated general instruction embeddings through self-supervised training on a large-scale unlabeled binary corpus, the efficiency of the PalmTree pre-trained language model is noted to be lower than that of traditional deep learning schemes like Instruction2Vec \cite{c21} which are not pre-trained. In the study of jTrans \cite{c21}, the researchers utilized a Transformer-based language model to embed control flow information of binary code, with a basic structure similar to BERT \cite{c8}, and trained it for the binary code similarity detection task. Moreover, the researchers introduced a new binary dataset called BinaryCorp \cite{c11}, which is currently the most diverse dataset. Notably, this paper employed the BinaryCorp dataset for the experiments. The authors of UniASM \cite{c12} introduced a novel pre-training model for analyzing assembly code, inspired by UniLM\cite{c4}. The model was trained on a self-constructed dataset and achieved promising results.

		\section{DESIGN OF FASTBCSD}
		\begin{figure*}[h]
		\centering
		\includegraphics[scale=0.36]{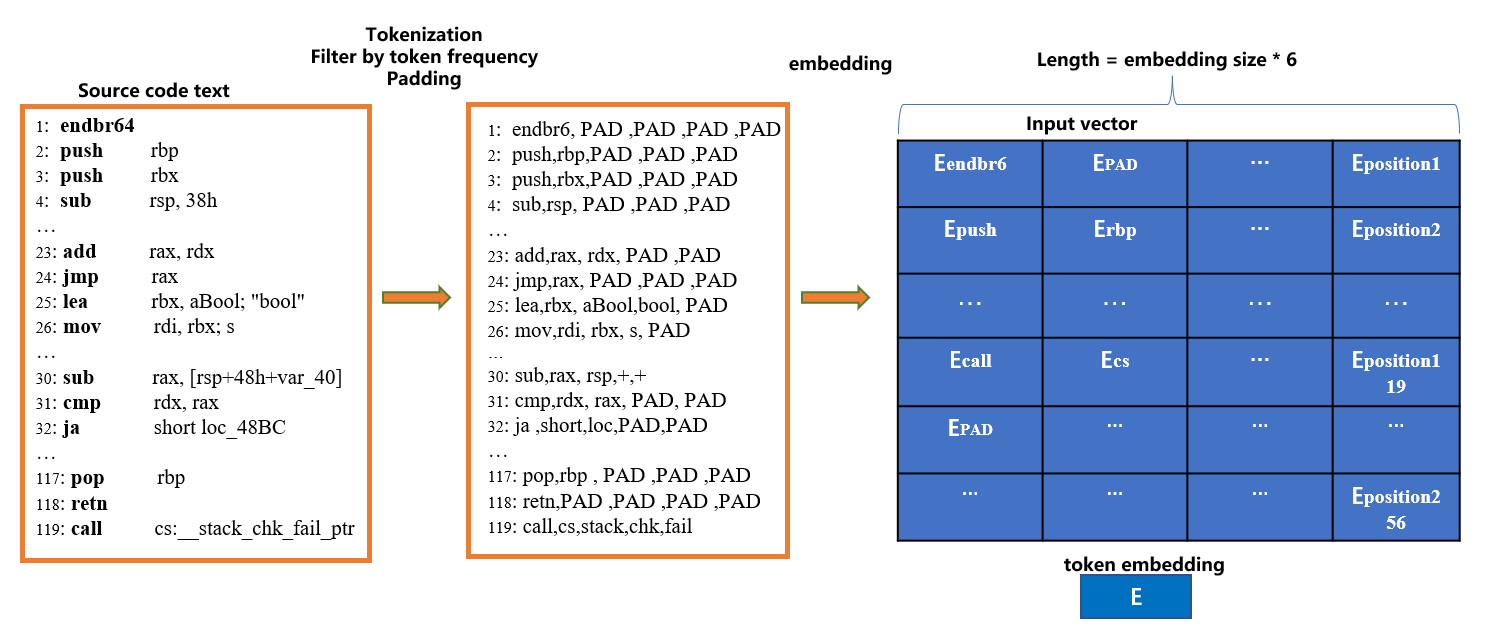}
		\caption{In the pre-processing stage of the code text, low-frequency tokens are removed and the number of tokens in each instruction is adjusted, following which the remaining tokens are converted into token embeddings. Subsequently, the token embeddings and a position embedding of each instruction are concatenated into an one-dimensional vector, serving as the instruction representation vector. All instruction vectors in the assembly code are concatenated into a two-dimensional vector, which is used as the input vector of the model.}
		\label{fig:my_label}
		\end{figure*}		
	
		\subsection{Text Preprocessing}
		Each assembly function consists of multiple instructions, each comprising an opcode and operands. While opcodes' usage in instructions is fixed, operand can vary due to the format characteristics of assembly language. In prior works, UniASM \cite{c12} treated the entire instruction as a single token, resulting in a high number of token types and the OOV(Out-Of-Vocabulary) problem. In prior studies, separating the opcode and operands from the instruction into distinct tokens was explored as a means of reducing the number of token types in assembly code analysis. To further mitigate the impact of varying instruction content on token types, normalization techniques were employed. For example, special tokens such as [str] were used in the PalmTree \cite{c10} and jTrans \cite{c11} studies to represent strings in instructions. In our study, we hypothesize that the OOV problem does not have a substantial impact on the effectiveness of our model. Rather than applying extensive normalization techniques to the operands, we have pursued a strategy of token mining, expanding the variety of tokens used in our analysis. Delimiters based on commonly used symbols in Instructions, such as "+", ":", ",", ";", and spaces, were utilized to decompose instructions into a multitude of tokens, resulting in hundreds of thousands of distinct tokens. We adopt a frequency-based filtering approach to select a subset of tokens from the large pool of tokens obtained by decomposing operands using various delimiters. Specifically, we increase the frequency of a token by one each time it appears in a training sample, and filter out tokens with a frequency less than a pre-defined hyperparameter F (in our final experiments, F is set to 32), resulting in a more manageable set of tokens with a size of approximately 40,000. The proposed approach preserves a greater quantity of semantic information through retaining more token types and more commonly occurring tokens. Figure \ref{fig:my_label} depicts the specific details of the preprocessing procedure.
		
		\subsection{Embedding Vector Construction}
		In the jTrans\cite{c11} study, researchers transformed the assembly text into a sequence of tokens by concatenating the opcode and operands in the order of their appearance within each instruction. The length of the token sequence is determined by the number of opcodes and operands in the assembly text. However, this approach presents a notable issue: as instructions in assembly language usually consist of an opcode and multiple operands, the number of parsed tokens can easily exceed 512 when the number of instructions surpasses 256. As jTrans employs the BERT architecture, it can only process token sequences with a maximum length of 512 tokens, leading to a discard of token sequences longer than 512, and in turn, resulting in poorer representation performance for lengthier assembly texts.

		In this paper, each assembly instruction is treated as a dynamic vector, and the length of the input sequence is determined by the number of instructions in the assembly code. The dynamic vector is obtained by concatenating an opcode token vector, multiple operand token vectors, and a positional vector along the feature dimension. To handle with the varying length of instructions, a threshold K is used to truncate instructions with token sequences longer than K, while special token vectors are used for padding when the number of tokens is less than K. In our experiments, we set the threshold K to 5. This value was chosen because the use of many symbols as separators in preprocessing can result in the parsing of many tokens for a single instruction. Therefore, a larger value of K is typically set to preserve more information for most instructions. In this approach, an assembly text composed of S instructions is represented as a two-dimensional vector of initialized embedding vectors with dimensions of S x H. Here, H denotes the dimensionality of a single token vector multiplied by K + 1. This vector is directly fed into TextCNN, LSTM, and MLP-Mixer for training. The assembled text vector can be conveniently utilized for training with TextCNN, LSTM, and MLP-Mixer models, as they don't pose any restriction on the input sequence length, in contrast to the Transformer model, which only supports input sequences of length less than 512. Figure \ref{fig:my_label} depicts the dimensionality of the instruction vector and the assembled text vector.
		
		\begin{figure*}[h]
		\centering
		\includegraphics[scale=0.32]{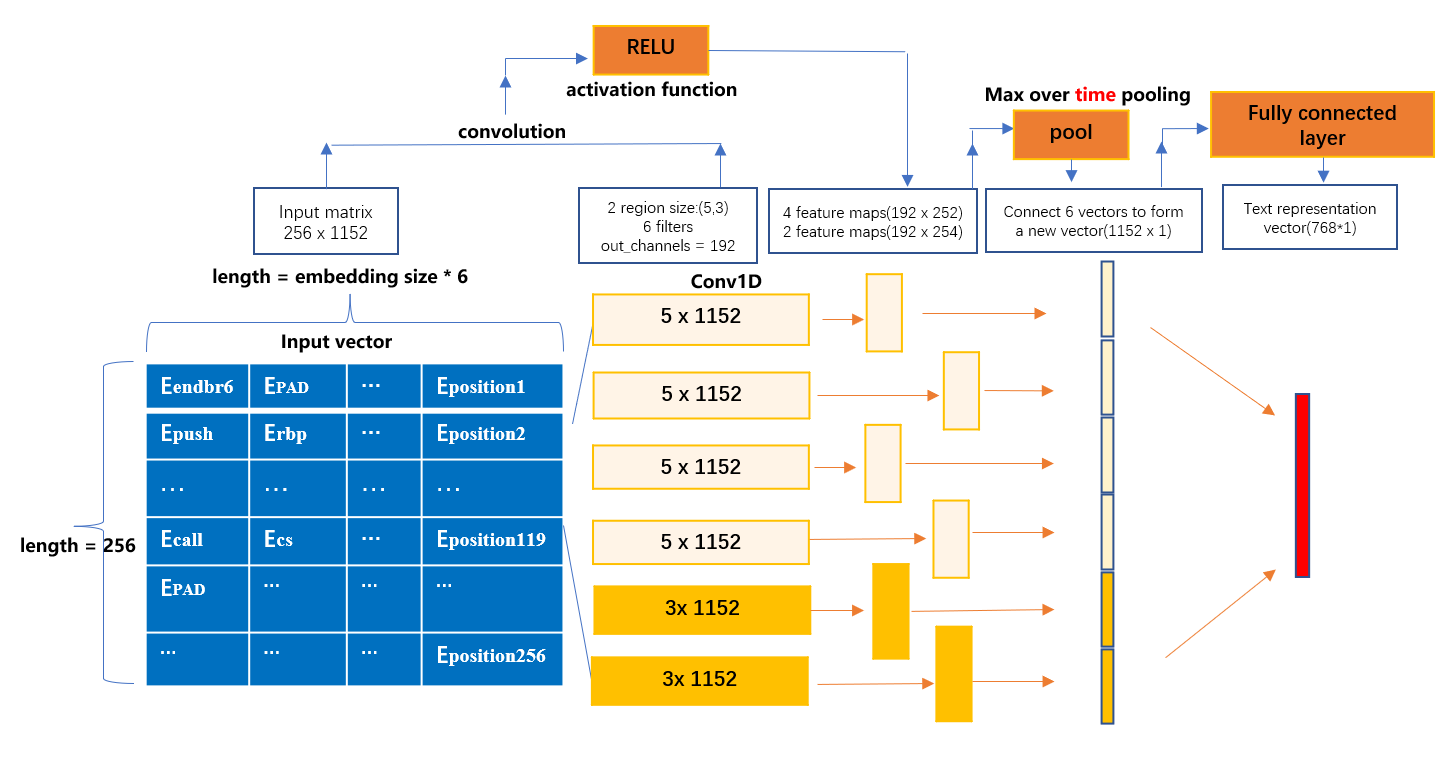}
		\caption{Extract text features using the TextCNN model with 6 one-dimensional convolutional kernels, a stride of 1, and an output channel of 192.}
		\label{fig2}
		\end{figure*}
		\subsection{Model Training}
		For training, a Siamese network framework \cite{c5} is utilized, with TextCNN serving as the model for feature extraction. As a type of text classification model, TextCNN is used in this study to extract the representation vector of assembly code. TextCNN follows a straightforward implementation approach. It receives a two-dimensional text vector (text length x vector dimension) as input and produces a text representation vector by sequentially passing through convolutional layers, activation functions, pooling layers, feature concatenation, and fully connected layers. The running process of the TextCNN model is presented in Figure \ref{fig2}. In addition to TextCNN, for comparison in feature extraction, we employed two other non-Transformer models, namely LSTM \cite{c15} and MLP-Mixer \cite{c19}. 
		
		LSTM (long-short term memory) is a type of recurrent neural network (RNN) model that is well-suited for handling and predicting important events in time series with long intervals and delays, thanks to its unique design structure. The application of LSTM for extracting the representation vector of assembly text has been reported in previous studies, where its input is a two-dimensional vector, such as the input vector shown in Figure 1. Its output is a one-dimensional vector containing global information, serving as the representation vector of assembly text. The calculation formula for the t-th time step of LSTM is as follows:
		\begin{equation}\label{eq:1}
		\mathrm{i}_{\mathrm{t}}=\sigma\left(\mathrm{W}_{\mathrm{i}} \cdot\left[\mathrm{h}_{\mathrm{t}-1}, \mathrm{x}_{\mathrm{t}}\right]+\mathrm{b}_{\mathrm{i}}\right)
		\end{equation}
		\begin{equation}\label{eq:2}
		\mathrm{f}_{\mathrm{t}}=\sigma\left(\mathrm{W}_{\mathrm{f}} \cdot\left[\mathrm{h}_{\mathrm{t}-1}, \mathrm{x}_{\mathrm{t}}\right]+\mathrm{b}_{\mathrm{f}}\right)
		\end{equation}
		\begin{equation}\label{eq:3}
		\tilde{\mathrm{C}}_{\mathrm{t}}=\tanh \left(\mathrm{W}_{\mathrm{C}} \cdot\left[\mathrm{h}_{\mathrm{t}-1}, \mathrm{x}_{\mathrm{t}}\right]+\mathrm{b}_{\mathrm{C}}\right)
		\end{equation}
		\begin{equation}\label{eq:4}
		\mathrm{o}_{\mathrm{t}}=\sigma\left(\mathrm{W}_{\mathrm{o}} \cdot\left[\mathrm{h}_{\mathrm{t}-1}, \mathrm{x}_{\mathrm{t}}\right]+\mathrm{b}_{\mathrm{o}}\right)
		\end{equation}
		\begin{equation}\label{eq:5}
		\mathrm{C}_{\mathrm{t}}=\mathrm{f}_{\mathrm{t}} * \mathrm{C}_{\mathrm{t}-1}+\mathrm{i}_{\mathrm{t}} * \tilde{\mathrm{C}}_{\mathrm{t}}
		\end{equation}
		\begin{equation}\label{eq:6}
		\mathrm{h}_{\mathrm{t}}=\mathrm{o}_{\mathrm{t}} * \tanh \left(\mathrm{C}_{\mathrm{t}}\right)
		\end{equation}
		The activation function sigmoid is denoted as $\sigma$, and $\tanh$ is also an activation function. $\mathrm{x}_{\mathrm{t}}$ represents the current input instruction vector, and $\mathrm{W}$ and $\mathrm{b}$ are learnable parameters. The output vector "ht" of the last time step serves as the representation vector of the assembly text.
	
\begin{figure*}[h]
			\centering
			\includegraphics[scale=0.27]{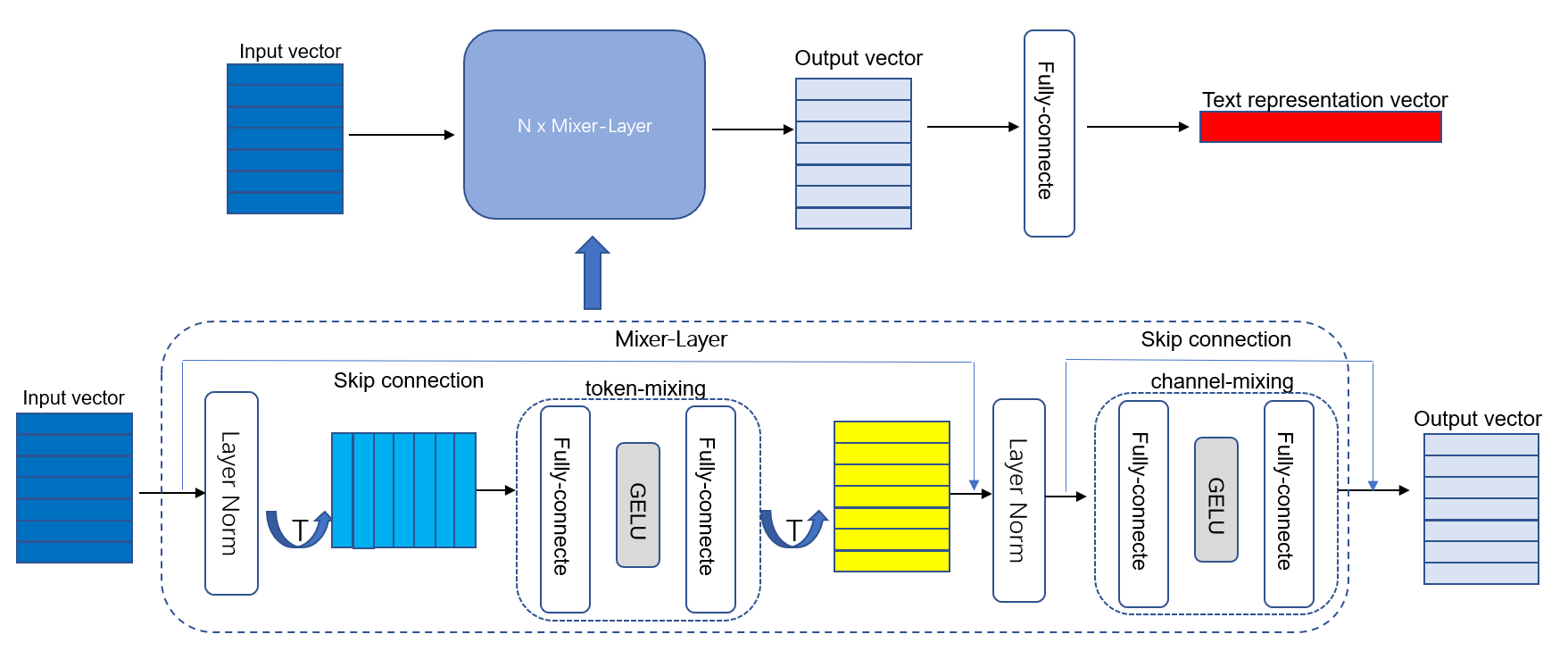}
			\caption{MLP-Mixer}
			\label{fig3}
		\end{figure*}

		The original MLP-Mixer consists of per-patch linear embeddings, Mixer layers, and a classifier head. In this paper, we remove the per-patch linear embeddings as our input is not image data and our assembly text vectors can be directly fed into the Mixer layers. The Mixer layers consist of one token-mixing MLP and one channel-mixing MLP, each containing a fully-connected layer and a GELU nonlinearity. The token-mixing MLP is used for feature mixing, while the channel-mixing MLP is used for token mixing, which can extract global information. Other components include skip-connections, dropout, and layer normalization.	
		The structure of the MLP-mixer used in this article is shown in Figure \ref{fig3}.
		
		 During the training process of the siamese network, a large number of positive and negative sample pairs are required. The label of the positive sample pair is 1, and the label of the negative sample pair is -1. The same TextCNN model is used to extract representation vectors for both functions in each sample pair, which are denoted as $E_{1}$ and $E_{2}$. After obtaining the representation vectors and labels, the cosine loss function is used to calculate the loss and update the network. The formula for the cosine loss function is as follows:
		{\small
		\begin{equation*}
		\min _{\theta} \mathcal{L}_{F}(\theta)=\left(\left(1-\cos \left(E_{1}, E_{2}\right)\right)+\max \left(0, \cos \left(E_{\mathrm{f}}, E_{\mathrm{g}}\right)-\operatorname{margin}\right)\right) * \frac{y+1}{2}
		\end{equation*}}
		where $\theta$ represents the parameters of the model, and margin is a hyper-parameter usually chosen between 0 and 1, we found that setting a relatively large value for the margin achieved the best results. In the experiments, we set it to 0.9.
		 
		\section{EXPERIMENTAL SETUP}
		\subsection{Dataset}
		\begin{table}[]
		\caption{Statistics on the number of projects, binaries and
			functions of the datasets.}
			\label{tab:t0}
			\begin{tabular}{cccc}
				\hline
				Datasets             & \# Projects & \# Binaries & \# Functions \\ \hline
				BinaryCorp-3M Train  & 1,612       & 8,357       & 3,126        \\ \hline
				BinaryCorp-3M Test   & 364         & 1,908       & 444,574      \\ \hline
				BinaryCorp-26M Train & 7,845       & 38,455      & 21,085,338   \\ \hline
				BinaryCorp-26M Test  & 1,974       & 9,675       & 4,791,673    \\ \hline
			\end{tabular}
		\end{table}
		This study used a publicly available large-scale binary dataset, BinaryCorp, which was first introduced in the jTrans\cite{c11} paper. BinaryCorp consists of a large number of binary documents, including official ArchLinux software packages and Arch user repositories, and 48,130 binary programs compiled with gcc and g++ at different optimization levels, with approximately 26 million functions in total. Due to the large size of the BinaryCorp-26M dataset, jTrans extracted a subset from it called BinaryCorp-3M. Table \ref{tab:t0} shows some statistics for these two datasets, with BinaryCorp-3M containing approximately 3.6 million functions. In this study, we used the training portion of BinaryCorp-3M as the training set for the entire model, and tested the trained model on the test sets of both BinaryCorp-3M and BinaryCorp-26M. It should be noted that the test set used in this study is the same as the one used in the jTrans article.
		\subsection{Data Sampling and Parameter Configuration}
		 In the BinaryCorp dataset, each original binary program is compiled at different optimization levels (O0, O1, O2, O3, OS) by the compiler, generating up to 5 functionally equivalent binary programs. We pair binary programs with the same functionality but different optimization levels to generate positive samples (filtering out positive samples with the same assembly text), and randomly sample R functionally different binary programs for each binary program to form negative samples. We found that the performance is better when R is around 30. We used this method to sample the training set of BinaryCorp-3M, generating about 47.6 million negative samples and 2.45 million positive samples, with a positive-to-negative ratio of approximately 1:19. The 50 million samples will be used as the final training samples. The dimension of both the word embedding and position embedding is 192. In TextCNN, we use four one-dimensional convolutional kernels with a size of 5 and two one-dimensional convolutional kernels with a size of 3. Each one-dimensional convolutional kernel has an input channel of 192*6 and an output channel of 192, with a stride of 1. The learning rate is set to 0.001, each batch contains 384 samples, and the model is trained for one epoch.
		\subsection{Evaluation Metrics}
		The application scenario of the BCSD model is to search for the function with the highest similarity to the input function from a large number of functions. Researchers usually use the MRR and Recall@k metrics to evaluate the performance of the model. In our experiments, we set a source function pool F, which has the same size as the number of functions, and the input function is selected from this pool. At the same time, we also set a target function pool G with the same size as the source function pool F. The definitions of the source function pool F and the target function pool G are as follows:
		\begin{equation}
		F=\left\{f_{1}, f_{2}, f_{3}, \ldots, f_{\mathrm{i}}, \ldots, f_{\mathrm{n}}\right\}
		\end{equation}
		\begin{equation}
		G=\left\{g_{1}, g_{2}, g_{3}, \ldots, g_{\mathrm{i}}, \ldots, 	g_{\mathrm{n}}\right\}
		\end{equation}		
		where $g_{1}$ $\in$ G and $f_{1}$ $\in$ F have the same functionality but different optimization levels, they form a positive sample pair. When calculating Mean Reciprocal Rank (MRR) and Recall@k, the similarity between $f_{1}$ and each function in G is first calculated, and then the functions in G are rearranged in descending order according to their similarity. $Rank_{g_{1}}$ denotes the position of $g_{1}$ in the reordered list. The formulas for calculating MRR and Recall@k are as follows:
		\begin{equation}
		M R R=\frac{1}{|F|} \sum_{f_{\mathrm{i}} \in F} \frac{1}{\left|\operatorname{Rank}_{g_{\mathrm{i}}}\right|}
		\end{equation}
		\begin{equation}		
		I(x)=\left\{\begin{array}{l}
		0, x>0 \\
		1, x \leq 0
		\end{array}\right.
		\end{equation}
		\begin{equation}		
		\operatorname{Recall@k}=\frac{1}{|F|} \sum_{f_{\mathrm{i}} \in F} \mathrm{I}\left(\operatorname{Rank}_{g_{\mathrm{i}}}-K\right)
		\end{equation}

{
	\small
	
	\begin{table*}[]
		\caption{Results of different binary similarity detection methods on BinaryCorp-3M (Poolsize=32)}
		\label{tab:t2}
		\begin{tabular}{cllllllllllllll}
			\hline
			\multicolumn{8}{c|}{MRR}                                                                                                                                                                                                                                                                                                   & \multicolumn{7}{c}{Recall@1}                                                                                                                                                                                                                                           \\ \hline
			\multicolumn{1}{c}{Models}                      & \multicolumn{1}{l}{O0,O3}          & \multicolumn{1}{l}{O1,O3}          & \multicolumn{1}{l}{O2,O3}          & \multicolumn{1}{l}{O0,Os}          & \multicolumn{1}{l}{O1,Os}          & \multicolumn{1}{l}{O2,Os}          & \multicolumn{1}{l}{Average}        & \multicolumn{1}{|l}{O0,O3}          & \multicolumn{1}{l}{O1,O3}          & \multicolumn{1}{l}{O2,O3}          & \multicolumn{1}{l}{O0,Os}          & \multicolumn{1}{l}{O1,Os}          & \multicolumn{1}{l}{O2,Os}          & \multicolumn{1}{l}{Average}        \\ \hline
			Gemini                                            & 0.388                               & 0.580                               & 0.750                               & 0.455                               & 0.546                               & 0.614                               & 0.556                               & \multicolumn{1}{|l}{0.238}                               & 0.457                               & 0.669                               & 0.302                               & 0.414                               & 0.450                               & 0.422                               \\
			SAFE                                              & 0.826                               & 0.917                               & 0.958                               & 0.854                               & 0.927                               & 0.927                               & 0.902                               & \multicolumn{1}{|l}{0.729}                               & 0.869                               & 0.933                               & 0.766                               & 0.879                               & 0.880                               & 0.843                               \\
			Asm2Vec                                           & 0.479                               & 0.878                               & 0.961                               & 0.536                               & 0.855                               & 0.900                               & 0.768                               & \multicolumn{1}{|l}{0.351}                               & 0.828                               & 0.942                               & 0.408                               & 0.796                               & 0.863                               & 0.701                               \\
			GraphEmb                                          & 0.602                               & 0.694                               & 0.750                               & 0.632                               & 0.674                               & 0.675                               & 0.671                               & \multicolumn{1}{|l}{0.485}                               & 0.600                               & 0.678                               & 0.521                               & 0.581                               & 0.584                               & 0.575                               \\
			OrderMatters-online                               & 0.542                               & 0.740                               & 0.869                               & 0.638                               & 0.702                               & 0.682                               & 0.695                               & \multicolumn{1}{|l}{0.414}                              & 0.647                               & 0.822                               & 0.515                               & 0.611                               & 0.593                               & 0.591                               \\
			OrderMatters                                      & 0.601                               & 0.838                               & 0.933                               & 0.701                               & 0.812                               & 0.800                               & 0.777                               & \multicolumn{1}{|l}{0.450}                               & 0.763                               & 0.905                               & 0.566                               & 0.724                               & 0.715                               & 0.687                               \\
			Genius                                            & 0.377                               & 0.587                               & 0.868                               & 0.437                               & 0.600                               & 0.627                               & 0.583                               & \multicolumn{1}{|l}{0.243}                               & 0.479                               & 0.830                               & 0.298                               & 0.490                               & 0.526                               & 0.478                               \\
			jTrans                                            & 0.947                               & 0.976                               & 0.985                               & 0.956                               & 0.979                               & 0.977                               & 0.970                               & \multicolumn{1}{|l}{0.913}                               & 0.960                               & 0.974                               & 0.927                               & 0.964                               & 0.961                               & 0.949                               \\ \hline
			\multicolumn{1}{c}{\textbf{FastBCSD-TextCNN}}   & \multicolumn{1}{l}{\textbf{0.931}} & \multicolumn{1}{l}{\textbf{0.971}} & \multicolumn{1}{l}{\textbf{0.981}} & \multicolumn{1}{l}{\textbf{0.945}} & \multicolumn{1}{l}{\textbf{0.976}} & \multicolumn{1}{l}{\textbf{0.970}} & \multicolumn{1}{l}{\textbf{0.962}} & \multicolumn{1}{|l}{\textbf{0.894}} & \multicolumn{1}{l}{\textbf{0.953}} & \multicolumn{1}{l}{\textbf{0.968}} & \multicolumn{1}{l}{\textbf{0.915}} & \multicolumn{1}{l}{\textbf{0.960}} & \multicolumn{1}{l}{\textbf{0.951}} & \multicolumn{1}{l}{\textbf{0.940}} \\ \hline
			\multicolumn{1}{c}{\textbf{FastBCSD-LSTM}}    & \textbf{0.909}                      & \textbf{0.964}                      & \textbf{0.977}                      & \textbf{0.932}                      & \textbf{0.971}                      & \textbf{0.961}                      & \textbf{0.952}                      & \multicolumn{1}{|l}{\textbf{0.864}}                      & \textbf{0.943}                     & \textbf{0.963}                      & \textbf{0.899}                      & \textbf{0.954}                      & \textbf{0.941}                      & \textbf{0.927}                     \\ \hline
			\multicolumn{1}{c}{\textbf{FastBCSD-MLP}} & \multicolumn{1}{l}{\textbf{0.900}} & \multicolumn{1}{l}{\textbf{0.964}} & \multicolumn{1}{l}{\textbf{0.978}} & \multicolumn{1}{l}{\textbf{0.921}} & \multicolumn{1}{l}{\textbf{0.969}} & \multicolumn{1}{l}{\textbf{0.961}} & \multicolumn{1}{l}{\textbf{0.948}} & \multicolumn{1}{|l}{\textbf{0.851}} & \multicolumn{1}{l}{\textbf{0.943}} & \multicolumn{1}{l}{\textbf{0.963}} & \multicolumn{1}{l}{\textbf{0.883}} & \multicolumn{1}{l}{\textbf{0.950}} & \multicolumn{1}{l}{\textbf{0.940}} & \multicolumn{1}{l}{\textbf{0.921}} \\ \hline
		\end{tabular}
	\end{table*}	
	\begin{table*}[]
		\caption{Results of different binary similarity detection methods on BinaryCorp-3M (Poolsize=10000)}
		\label{tab:t3}
		\begin{tabular}{cllllllllllllll}
			\hline
			\multicolumn{8}{c}{MRR}                                                                                                                                                                                                                                                                                                   & \multicolumn{7}{c}{Recall@1}                                                                                                                                                                                                                                           \\ \hline
			\multicolumn{1}{c}{Models}                      & \multicolumn{1}{l}{O0,O3}          & \multicolumn{1}{l}{O1,O3}          & \multicolumn{1}{l}{O2,O3}          & \multicolumn{1}{l}{O0,Os}          & \multicolumn{1}{l}{O1,Os}          & \multicolumn{1}{l}{O2,Os}          & \multicolumn{1}{l}{Average}        & \multicolumn{1}{|l}{O0,O3}          & \multicolumn{1}{l}{O1,O3}          & \multicolumn{1}{l}{O2,O3}          & \multicolumn{1}{l}{O0,Os}          & \multicolumn{1}{l}{O1,Os}          & \multicolumn{1}{l}{O2,Os}          & \multicolumn{1}{l}{Average}        \\ \hline
			Gemini                                            & 0.037                               & 0.161                               & 0.416                               & 0.049                               & 0.133                               & 0.195                               & 0.165                               & \multicolumn{1}{|l}{0.024}                               & 0.122                               & 0.367                               & 0.030                               & 0.099                               & 0.151                               & 0.132                               \\
			SAFE                                              & 0.127                               & 0.345                               & 0.643                               & 0.147                               & 0.321                               & 0.377                               & 0.320                               & \multicolumn{1}{|l}{0.068}                              & 0.247                               & 0.575                               & 0.079                               & 0.221                               & 0.283                               & 0.246                               \\
			Asm2Vec                                           & 0.072                               & 0.449                               & 0.669                               & 0.083                               & 0.409                               & 0.510                               & 0.366                               & \multicolumn{1}{|l}{0.046}                               & 0.367                               & 0.589                               & 0.052                               & 0.332                               & 0.426                               & 0.302                               \\
			GraphEmb                                          & 0.087                               & 0.217                               & 0.486                               & 0.110                               & 0.195                               & 0.222                               & 0.219                               & \multicolumn{1}{|l}{0.050}                               & 0.154                               & 0.447                               & 0.063                               & 0.135                               & 0.166                               & 0.169                               \\
			OrderMatters                                      & 0.062                               & 0.319                               & 0.600                               & 0.075                               & 0.260                               & 0.233                               & 0.263                               & \multicolumn{1}{|l}{0.040}                               & 0.248                               & 0.535                               & 0.040                               & 0.178                               & 0.158                               & 0.200                               \\
			Genius                                            & 0.041                               & 0.193                               & 0.596                               & 0.049                               & 0.186                               & 0.224                               & 0.214                               & \multicolumn{1}{|l}{0.028}                                & 0.153                               & 0.538                               & 0.032                               & 0.146                               & 0.180                               & 0.179                               \\
			jTrans                                            & 0.475                               & 0.663                               & 0.731                               & 0.539                               & 0.665                               & 0.664                               & 0.623                               & \multicolumn{1}{|l}{0.376}                               & 0.580                               & 0.661                               & 0.443                               & 0.586                               & 0.585                               & 0.538                               \\ \hline
			\multicolumn{1}{c}{\textbf{FastBCSD-TextCNN}}   & \multicolumn{1}{l}{\textbf{0.485}} & \multicolumn{1}{l}{\textbf{0.662}} & \multicolumn{1}{l}{\textbf{0.742}} & \multicolumn{1}{l}{\textbf{0.558}} & \multicolumn{1}{l}{\textbf{0.681}} & \multicolumn{1}{l}{\textbf{0.679}} & \multicolumn{1}{l}{\textbf{0.633}} & \multicolumn{1}{|l}{\textbf{0.389}} & \multicolumn{1}{l}{\textbf{0.577}} & \multicolumn{1}{l}{\textbf{0.675}} & \multicolumn{1}{l}{\textbf{0.461}} & \multicolumn{1}{l}{\textbf{0.599}} & \multicolumn{1}{l}{\textbf{0.600}} & \multicolumn{1}{l}{\textbf{0.550}} \\ \hline
			\multicolumn{1}{c}{\textbf{FastBCSD-LSTM}}      & \multicolumn{1}{l}{\textbf{0.437}} & \multicolumn{1}{l}{\textbf{0.645}} & \multicolumn{1}{l}{\textbf{0.727}} & \multicolumn{1}{l}{\textbf{0.530}} & \multicolumn{1}{l}{\textbf{0.667}} & \multicolumn{1}{l}{\textbf{0.653}} & \multicolumn{1}{l}{\textbf{0.610}} & \multicolumn{1}{|l}{\textbf{0.349}} & \multicolumn{1}{l}{\textbf{0.563}} & \multicolumn{1}{l}{\textbf{0.659}} & \multicolumn{1}{l}{\textbf{0.441}} & \multicolumn{1}{l}{\textbf{0.585}} & \multicolumn{1}{l}{\textbf{0.573}} & \multicolumn{1}{l}{\textbf{0.528}} \\ \hline
			\multicolumn{1}{c}{\textbf{FastBCSD-MLP}} & \multicolumn{1}{l}{\textbf{0.398}} & \multicolumn{1}{l}{\textbf{0.638}} & \multicolumn{1}{l}{\textbf{0.721}} & \multicolumn{1}{l}{\textbf{0.476}} & \multicolumn{1}{l}{\textbf{0.660}} & \multicolumn{1}{l}{\textbf{0.652}} & \multicolumn{1}{l}{\textbf{0.590}} & \multicolumn{1}{|l}{\textbf{0.309}} & \multicolumn{1}{l}{\textbf{0.557}} & \multicolumn{1}{l}{\textbf{0.652}} & \multicolumn{1}{l}{\textbf{0.387}} & \multicolumn{1}{l}{\textbf{0.582}} & \multicolumn{1}{l}{\textbf{0.574}} & \multicolumn{1}{l}{\textbf{0.510}} \\ \hline
		\end{tabular}
	\end{table*}	
	\begin{table*}[]
		\caption{Results of different binary similarity detection methods on BinaryCorp-26M (Poolsize=32).It should be noted that the FastBCSD model uses the training set from BinaryCorp-3M, while the other models in the table employ BinaryCorp-26M, which is a superset of BinaryCorp-3M.}
		\label{tab:t4}	
		\begin{tabular}{ccccccccccccccc}
			\hline
			\multicolumn{8}{c|}{MRR}                                                                                                                                                                                                                                                                                                 & \multicolumn{7}{c}{Recall@1}                                                                                                                                                                                                                                            \\ \hline
			\multicolumn{1}{c}{Models}                    & \multicolumn{1}{c}{O0,O3}          & \multicolumn{1}{c}{O1,O3}          & \multicolumn{1}{c}{O2,O3}          & \multicolumn{1}{c}{O0,Os}          & \multicolumn{1}{c}{O1,Os}          & \multicolumn{1}{c}{O2,Os}          & \multicolumn{1}{c|}{Average}        & \multicolumn{1}{c}{O0,O3}          & \multicolumn{1}{c}{O1,O3}          & \multicolumn{1}{c}{O2,O3}          & \multicolumn{1}{c}{O0,Os}          & \multicolumn{1}{c}{O1,Os}          & \multicolumn{1}{c}{O2,Os}          & \multicolumn{1}{c}{Average}         \\ \hline
			Gemini                                          & 0.402                               & 0.643                               & 0.835                               & 0.469                               & 0.564                               & 0.628                               & 0.590                               & \multicolumn{1}{|l}{0.263}                               & 0.528                               & 0.768                               & 0.322                               & 0.441                               & 0.518                               & 0.473                                \\
			SAFE                                            & 0.856                               & 0.940                               & 0.970                               & 0.874                               & 0.935                               & 0.934                               & 0.918                               & \multicolumn{1}{|l}{0.770}                               & 0.902                               & 0.951                               & 0.795                               & 0.891                               & 0.891                               & 0.867                                \\
			Asm2Vec                                         & 0.439                               & 0.847                               & 0.958                               & 0.490                               & 0.788                               & 0.849                               & 0.729                               & \multicolumn{1}{|l}{0.314}                               & 0.789                               & 0.940                               & 0.362                               & 0.716                               & 0.800                               & 0.654                                \\
			GraphEmb                                        & 0.583                               & 0.681                               & 0.741                               & 0.610                               & 0.637                               & 0.639                               & 0.649                               & \multicolumn{1}{|l}{0.465}                               & 0.586                               & 0.667                               & 0.499                               & 0.541                               & 0.543                               & 0.550                                \\
			OrderMatters                                    & 0.572                               & 0.820                               & 0.932                               & 0.630                               & 0.692                               & 0.771                               & 0.729                               & \multicolumn{1}{|l}{0.417}                               & 0.740                               & 0.903                               & 0.481                               & 0.692                               & 0.677                               & 0.652                                \\
			jTrans                                          & 0.964                               & 0.983                               & 0.989                               & 0.969                               & 0.980                               & 0.980                               & 0.978                               & \multicolumn{1}{|l}{0.941}                               & 0.970                               & 0.981                               & 0.949                               & 0.964                               & 0.964                               & 0.962                                \\ \hline
			\multicolumn{1}{c}{\textbf{FastBCSD-TextCNN}} & \multicolumn{1}{c}{\textbf{0.926}} & \multicolumn{1}{c}{\textbf{0.968}} & \multicolumn{1}{c}{\textbf{0.976}} & \multicolumn{1}{c}{\textbf{0.944}} & \multicolumn{1}{c}{\textbf{0.972}} & \multicolumn{1}{c}{\textbf{0.967}} & \multicolumn{1}{c}{\textbf{0.959}} & \multicolumn{1}{|c}{\textbf{0.888}} & \multicolumn{1}{c}{\textbf{0.949}} & \multicolumn{1}{c}{\textbf{0.963}} & \multicolumn{1}{c}{\textbf{0.916}} & \multicolumn{1}{c}{\textbf{0.955}} & \multicolumn{1}{c}{\textbf{0.948}} & \multicolumn{1}{c}{\textbf{0.9368}} \\ \hline
		\end{tabular}

	\end{table*}                                                                            
	\begin{table*}[]
		\caption{Results of different binary similarity detection methods on BinaryCorp-26M (Poolsize=10000).It should be noted that the FastBCSD model uses the training set from BinaryCorp-3M, while the other models in the table employ BinaryCorp-26M, which is a superset of BinaryCorp-3M. }
		\label{tab:t5}	
		\begin{tabular}{cllllllllllllll}
			\hline
			\multicolumn{8}{c|}{MRR}                                                                                                                                                                                                                                                                                                 & \multicolumn{7}{c}{Recall@1}                                                                                                                                                                                                                                           \\ \hline
			\multicolumn{1}{c}{Models}                    & \multicolumn{1}{l}{O0,O3}          & \multicolumn{1}{l}{O1,O3}          & \multicolumn{1}{l}{O2,O3}          & \multicolumn{1}{l}{O0,Os}          & \multicolumn{1}{l}{O1,Os}          & \multicolumn{1}{l}{O2,Os}          & \multicolumn{1}{l}{Average}        & \multicolumn{1}{|l}{O0,O3}          & \multicolumn{1}{l}{O1,O3}          & \multicolumn{1}{l}{O2,O3}          & \multicolumn{1}{l}{O0,Os}          & \multicolumn{1}{l}{O1,Os}          & \multicolumn{1}{l}{O2,Os}          & \multicolumn{1}{l}{Average}        \\ \hline
			Gemini                                          & 0.072                               & 0.189                               & 0.474                               & 0.069                               & 0.147                               & 0.202                               & 0.192                               & \multicolumn{1}{|l}{0.058}                               & 0.148                               & 0.420                               & 0.051                               & 0.115                               & 0.162                               & 0.159                               \\
			SAFE                                            & 0.198                               & 0.415                               & 0.696                               & 0.197                               & 0.377                               & 0.431                               & 0.386                               & \multicolumn{1}{|l}{0.135}                               & 0.314                               & 0.634                               & 0.127                               & 0.279                               & 0.343                               & 0.305                               \\
			Asm2Vec                                         & 0.118                               & 0.443                               & 0.703                               & 0.107                               & 0.369                               & 0.480                               & 0.370                               & \multicolumn{1}{|l}{0.099}                               & 0.376                               & 0.638                               & 0.086                               & 0.307                               & 0.413                               & 0.320                               \\
			GraphEmb                                        & 0.116                               & 0.228                               & 0.498                               & 0.133                               & 0.198                               & 0.224                               & 0.233                               & \multicolumn{1}{|l}{0.080}                               & 0.171                               & 0.465                               & 0.090                               & 0.145                               & 0.175                               & 0.188                               \\
			OrderMatters                                    & 0.113                               & 0.292                               & 0.682                               & 0.118                               & 0.256                               & 0.295                               & 0.292                               & \multicolumn{1}{|l}{0.094}                               & 0.222                               & 0.622                               & 0.093                               & 0.195                               & 0.236                               & 0.244                               \\
			jTrans                                          & 0.584                               & 0.734                               & 0.792                               & 0.627                               & 0.709                               & 0.710                               & 0.693                               & \multicolumn{1}{|l}{0.499}                               & 0.668                               & 0.736                               & 0.550                               & 0.648                               & 0.648                               & 0.625                               \\ \hline
			\multicolumn{1}{c}{\textbf{FastBCSD-TextCNN}} & \multicolumn{1}{l}{\textbf{0.535}} & \multicolumn{1}{l}{\textbf{0.710}} & \multicolumn{1}{l}{\textbf{0.772}} & \multicolumn{1}{l}{\textbf{0.607}} & \multicolumn{1}{l}{\textbf{0.713}} & \multicolumn{1}{l}{\textbf{0.709}} & \multicolumn{1}{l}{\textbf{0.674}} & \multicolumn{1}{|l}{\textbf{0.456}} & \multicolumn{1}{l}{\textbf{0.643}} & \multicolumn{1}{l}{\textbf{0.717}} & \multicolumn{1}{l}{\textbf{0.527}} & \multicolumn{1}{l}{\textbf{0.647}} & \multicolumn{1}{l}{\textbf{0.646}} & \multicolumn{1}{l}{\textbf{0.606}} \\ \hline
		\end{tabular}
	\end{table*}
	
}

		\section{EVALUATION}
		In order to establish a standardized evaluation framework, we employed the publicly available code of jTrans to construct the testing datasets, BinaryCorp-3M and BinaryCorp-26M. Furthermore, we applied our pre-processing and tokenization techniques to prepare the functions in the testing datasets. The model training and inference was conducted on a hardware environment consisting of an Intel Xeon 10-core 2.20GHz CPU, 256GB RAM, and 1 Nvidia 1080Ti GPU.
		\subsection{Biniary Similarity Detection Performance}
		For the test sets of BinaryCorp-3M and BinaryCorp-26M, we adopted the same function pool sizes as jTrans\cite{c11}, which are 32 and 10000, respectively. Our experimental results can be seen in Tables\ref{tab:t2}-\ref{tab:t5} (except for FastBCSD, the performance data of other models are from jTrans' experimental results. Table \ref{tab:t2} reports the recall@1 scores of several BCSD models, including jTrans, on BinaryCorp-3M and BinaryCorp-26M test sets. We noticed a minor error in the recall@1 score of jTrans reported in the original table, which did not significantly affect the results. However, we recalculated the value based on the provided numerical information and obtained a revised score of 0.538. Our FastBCSD-TextCNN method achieves similar performance to jTrans and outperforms other BCSD models significantly. For the test task with a function pool size of 32, which is easier but significantly different from real-world scenarios, our proposed FastBCSD-TextCNN method achieves an MRR value that is slightly lower than that of jTrans, with a difference of 0.01 to 0.02. However, when the function pool size is 10000, which is closer to real-world scenarios, FastBCSD-TextCNN outperforms the best-performing jTrans model on BinaryCorp-3M with a difference of 0.01 in MRR value and 0.012 in recall@1 value. When evaluating on the larger BinaryCorp-26M dataset, FastBCSD-TextCNN achieved an MRR value that is 0.02 lower and a recall@1 value that is 0.02 lower than the jTrans model. The performance gap between the two models may be attributed to the fact that we only used the training set of BinaryCorp-3M and did not utilize the training set of BinaryCorp-26M due to its large size. Based on the experimental findings, the performance of FastBCSD-TextCNN is similar to that of jTrans, yet it shows significant improvement over other baseline models in terms of MRR and recall values, except for jTrans. Particularly, when the function pool size is 10000, the MRR value of FastBCSD-TextCNN is higher by 0.31 compared to SAFE, a model constructed with the Siamese network framework and bidirectional LSTM. Based on the results shown in Figure \ref{tab:t9}, it can be observed that the parameter size of FastBCSD-TextCNN is only 15$\%$ of that of jTrans, and its inference time is only 1/5 of jTrans. Meanwhile, the inference time of FastBCSD-MLP is even less, about 1/7 of jTrans. In contrast, the slowest model is FastBCSD-LSTM, which is due to the recurrent neural network architecture used.

		The performance comparison of different models is presented in Tables \ref{tab:t2}-\ref{tab:t3}. The results demonstrate that FastBCSD-TextCNN outperforms all other models, while FastBCSD-MLP based on MLP-Mixer exhibits the worst performance. Nevertheless, it is observed that even the worst-performing model, FastBCSD-MLP, is not significantly behind jTrans. Table \ref{tab:t3} shows that FastBCSD-MLP achieves an MRR score that is only 0.033 lower than jTrans, while outperforming the SAFE model by 0.27. The SAFE model is based on the bidirectional LSTM plus the Siamese network framework, which is similar to the model used in FastBCSD-LSTM. However, the MRR score of FastBCSD-LSTM is significantly higher than that of SAFE, indicating that our dynamic instruction scheme can effectively adapt to smaller models.

		\subsection{Reflection on experimental results}
		One important factor contributing to the success of small models in our study is the ability to construct a large amount of supervised data in batches for the BCSD task. This allows for a substantial volume of data to be generated and a relatively uniform data distribution. The proposed approach in this study, which trains small models using a large amount of supervised data, can achieve comparable performance to the pre-training and fine-tuning approach that uses large models. It should be noted that not all small model approaches are able to achieve comparable results to pre-trained models when trained on a large amount of supervised data, as these approaches may not take into account the impact of data volume on performance. Many studies rely on small-scale training data, which is susceptible to overfitting. The present study uses the largest open dataset available, which helps to avoid the overfitting problem. The popularity of pre-training approaches in NLP is well-known, mainly because most NLP tasks are not as well-defined as BCSD tasks, and acquiring a sufficient amount of supervised data for NLP tasks is often difficult.

	\section{DISCUSSION}
	In this article, our research is focused solely on the x86 instruction set. But FastBCSD could be applied to other types of assembly languages such as ARM and MIPS. However, we have not conducted experiments on cross-platform binary function recognition in this study. This task is more challenging since the representation of the same function varies significantly across different instruction sets, and there are additional issues such as semantic alignment between different assembly languages. In the future, we plan to make modifications to FastBCSD to adapt it to the task of cross-platform binary function recognition.
	
	\section{CONCLUSION}
	In this paper, we propose a novel dynamic instruction vector encoding method taking only assembly code as input features combing with a lightweight neural network, named FastBCSD, to address some limitations of current methods in BCSD task, such as the training of pre-trained models requires a significant amount of computational resources, and normalization of tokens can lead to loss of information. Experimental results show that FastBCSD achieves a similar performance with the state-of-the-art pre-training model, but with significantly fewer parameters and faster inference speed.
	
\begin{table}[]
	\caption{The inference time and model parameter size vary across different models. The inference time refers to the average time required to infer each sample in the test set, measured in seconds.}
	\label{tab:t9}
	\begin{tabular}{|c|c|c|}
		\hline
		{\color[HTML]{000000} model} & {\color[HTML]{000000} inference time} & {\color[HTML]{000000} model size} \\ \hline
		jTrans                       & 0.0129s                               & 87.88m                            \\ \hline
		FastBCSD-TextCNN             & 0.00267s                              & 13.44m                            \\ \hline
		FastBCSD-LSTM                & 0.00371s                              & 9.77m                             \\ \hline
		FastBCSD-MLP                 & 0.00195s                              & 13.04m                            \\ \hline
	\end{tabular}
	
\end{table}

	\bibliographystyle{ACM-Reference-Format}
	\bibliography{sample-sigconf}

	\appendix

\end{document}